# Observations of temporal group delays in slow-light multiple coupled photonic crystal cavities

S. Kocaman[1*], X. Yang[2], J. F. McMillan[1], M. B.Yu[3], D. L. Kwong[3], and C. W. Wong[1*]

[1]*Optical Nanostructures Laboratory, Center for Integrated Science and Engineering, Solid-State Science and Engineering, and Mechanical Engineering, Columbia University, New York, NY 10027, USA*

[2]*University of California at Berkeley and Lawrence Berkeley National Laboratory, Berkeley, CA 94720, USA*

[3]*The Institute of Microelectronics, 11 Science Park Road, Singapore Science Park II Singapore 117685, Singapore*

We demonstrate temporal group delays in coherently-coupled high-$Q$ multi-cavity photonic crystals, in an all-optical analogue to electromagnetically induced transparency. We report deterministic control of the group delay up to 4× the single cavity lifetime in our CMOS-fabricated room-temperature chip. Supported by three-dimensional numerical simulations and theoretical analyses, our multi-pump beam approach enables control of the multi-cavity resonances and inter-cavity phase, in both single and double transparency peaks. The standing-wave wavelength-scale photon localization allows direct scalability for chip-scale optical pulse trapping and coupled-cavity QED.

PACS numbers: 42.60.Da, 42.50.Gy, 42.70.Qs, 42.82.-m

* Authors to whom correspondence should be addressed. Electronic addresses: sk2927@columbia.edu ; cww2104@columbia.edu .

Large-scale communication systems have benefited from using photons as the transport medium, such as the optical fiber network infrastructure and massively parallel computing architectures. A subset of these efforts is the recent examination of optical interconnects and optical delay lines, where chip-scale implementations can provide scalability, integration, low-power dissipation and large bandwidths. Actively tunable delay lines provide a step towards network or communications reconfigurability. In electromagnetically induced transparency (EIT) – a remarkable outcome of the quantum coherence in atoms – destructive quantum interference introduced by a strong coupling laser cancels the ground state absorption to coherent superposing upper states in a three-level system [1]. This observation of sharp cancellation of absorption resonance through atomic coherence has led to phenomena such as lasing without inversion [2], frozen light using the steep linear dispersion from extremely narrow linewidth of the EIT [3], and dynamic storage of light greater than a second in solid-state materials [4].

Several theoretical analyses of coupled optical resonators has revealed that coupled resonator systems can have interestingly similar phenomena to atomic systems where the interference of resonant pathways with resulting EIT-like spectra is enforced by the geometry or dispersion of nanophotonic structures [5-8]. In this all-optical classical analog, a single or multiple sharp transparency windows can be induced by coherent interferences between normal modes in coupled optical resonators in an originally non-transmitting background with transparency linewidths at GHz or more, significantly broader than the narrow (~ 100 kHz or less) linewidths in atomic systems. These dispersive slow-light effects [9-10] were recently examined experimentally in two coupled whispering-gallery mode resonators [11], ultrahigh-$Q$ microspheres [12], and two or more coupled photonic crystal cavities [13], with applications in trapping light at room temperature beyond the fundamental delay-bandwidth product [5-8]. Here we report the experimental time-domain observations of delays up to 17.12 ps, or more than 4× the single cavity lifetime, in the near-infrared.

Our system, illustrated in Fig. 1a, consists of a photonic crystal waveguide side-coupled to four photonic crystal cavities. As recently demonstrated in multi-EIT-like lineshapes [13], we used defect-type cavities, formed with three missing air holes (*L3*) in an air-bridged hexagonal lattice photonic crystal membrane. From coupled mode theory [14], the dynamical equations for the cavity mode amplitudes are

$$\frac{da_n}{dt} = \left(-\frac{1}{2\tau_{total,n}} + i(\omega_1 + \Delta\omega_1 - \omega_{wg})\right)a_n + \kappa s_{R(n-1)} + \kappa s_{Ln}$$

where $n$ is the cavity number, $\omega$ is the resonant frequency, $a$ is the normalized cavity mode amplitude and $s$ is the normalized waveguide mode amplitude. Without significant nonlinear absorption [15], the total loss rate for the resonance mode is described by $\frac{1}{\tau_{total}} = \frac{1}{\tau_v} + \frac{1}{\tau_{in}}$, where $\frac{1}{\tau_{v(in)}}$ ($=\frac{\omega}{Q_{v(in)}}$) is the decay rate from the cavity into the continuum (waveguide) and $\kappa$ ($= ie^{(-i\phi/2)}/\sqrt{2\tau_{in}}$) is the cavity-waveguide coupling rate. Phase between the cavities is given by $\phi = \omega_{wg} n_{eff} L/c$ where $n_{eff}$ is 2.768 at 1550 nm. These parameters are computed numerically using 3D finite difference time domain (FDTD) calculations (Fig. 1c) and the samples are fabricated using 248 nm UV lithography, with low (sub-20 Å) statistically quantified disorder [16] (example scanning electron microscopy (SEM) shown in Fig. 1d-e). The single-crystal device layer is 250 nm thick and on top of a 1 μm buried oxide; the buried oxide is subsequently wet-etched away to form an air-bridged membrane structure for our experiments. The membrane has a thickness of 0.595$a$ and hole radius of 0.261$a$, where the lattice period $a$ = 420 nm. The position of the nearest neighbor holes are shifted by 0.15$a$ to tune the radiation mode field for increasing the intrinsic $Q$ factors [17]. The separation between the center of the waveguide and the center of the cavity is 2√3$a$. For a single cavity, the $Q_{tot}$ was determined experimentally to be 5,110 and, using the measured $Q_{tot}$ and the measured -13.17dB resonance intensity contrast ratio, we determined $Q_v$ and $Q_{in}$ to be at 23,261 and 6,548 respectively [18]. The experimental $Q_v$ factor is different than designed $Q_v$ of 60,000 mainly because of the fluctuations in the fabricated structures, small perturbations and slightly angled etched sidewalls. The modal volume $V$ is ~ 0.74 cubic wavelengths [$(\lambda/n)^3$]. Large $Q_v/Q_{in}$ ratios are needed to be high in order to operate in the overcoupled regime for strong in-plane interference. The interacting multiple cavities are designed identically; however due the slight fabrication deviations, the actual resonant frequencies are not exactly the same but with a slight detuning which allows us to obtain an EIT-like transparency peak. When the two cavity resonances are close enough and the cavity-to-cavity round trip phase satisfy the condition of forming a Fabry-Pérot resonance (2$n\pi$), the system represents an all-optical analogue of EIT which has been observed spectrally with two and three cavities in our recent lineshape studies [13], resulting a photon delay that is longer than both cavity lifetimes (calculated as 4.15 ps each) combined.

To perform the group velocity delay measurements, we build the experimental setup is shown in Fig. 2, where a 1 GHz modulated tunable laser is coupled into the chip-scale multi-cavity system. In order to increase the coupling efficiency between the optical fiber and the waveguides [19], we used nanotapered structures at both the input and output waveguides. Two polarization controllers are used; first one is for getting the optimal modulation and the second one for tuning the light to transverse-electric (TE) polarization and then couple into the chip with a tapered lensed fiber. At the waveguide output, another tapered lensed fiber is used to collect the output signal. On the output side, an erbium doped fiber amplifier (EDFA) is used to get the desired power level for the high speed photo detector and an automatic channel locking filter (Digital Lightwave) is used to filter the EDFA noise. Finally, a digital sampling oscilloscope recorded the relative delays between different wavelengths, synchronized to the input modulator with 1 ps accuracy [20]. To align the cavity resonances and tune the phase between cavities, we thermo-optically tune the chip locally with two frequency-doubled 532-nm pump lasers to get the desired state by increasing refractive index of silicon with a rate of 1.85 $\times 10^{-4}$ /K at 1.55 µm [15,21]. We estimate 16 K temperature rise per milliwatt pump (~1.32nm/mW resonance redshift and 0.0153π/mW phaseshift).

Here we present results from three series of experiments. Fig. 3 shows the comparison between measured and calculated transmission spectra, $T = |s_{R4}|^2 / |s_{R0}|^2$, and the corresponding measured and calculated group delay values, $d\phi/d\omega$. Our system consist of a four L3 cavities coupled to a single line-defect photonic crystal waveguide (as shown in Fig.1d), where we work with three of the cavity resonances with closest frequency spacing. Fig. 3a shows the transmission spectrum of three cavities (without external tuning) where two of them are almost at the same frequency, with the cavity resonances are $\lambda_1$ = 1533.52 nm, $\lambda_2$ = 1533.98 nm and $\lambda_3$ = 1534.02 nm – the slight (3.3%) resonant frequency difference is due to fabrication deviations between cavities. Note that in this spectrum there is no transparency peaks due to the large frequency detuning between cavity 1 and cavities 2-3, and a phase mismatch between cavity 2 and cavity 3.

Correspondingly, in the temporal delay measurements (Fig. 3e), there is no significant spectral feature as expected, except for the Fabry-Perot-type reflection noise from the chip end facets.

Fig. 3b shows the transmission spectrum with external tuning, with the resonances deterministically tuned to longer wavelengths. By adjusting the spatial position of the pump, we shift two resonances close enough for the EIT-like detuning condition $\delta_{23} <$ ~3.5 [13]. In addition, we focus the second pump laser on the photonic crystal waveguide between the cavities in order to get the required phase condition ($\phi = n\pi$, where $n$ is an integer). As shown in Fig. 3b, the coherent transparency peak appears at 1534.26 nm ($\delta_{23}$ = 0.79), with the resonant frequencies now at $\lambda_1$ = 1534.10 nm, $\lambda_2$ = 1534.20 nm and $\lambda_3$ = 1534.32 nm. A larger shift at $\lambda_1$ occurred due to the local heating. We note there is only one transparency peak due to the phase mismatch between cavity 1 and 2. Fig. 3f now shows the corresponding temporal delay values. We observed a 17.12 ps delay at the transparency wavelength, or equivalent to 4× of the single cavity lifetimes (4.15ps) and more than two incoherently summed cavity lifetimes. The slow-down factor $S$, or the ratio of the phase velocity to the group velocity ($v_\phi/v_g$), is determined to be ~150 at the transparency peak [10]. We match the spectral features of both the transmission and the delay spectrum by breaking the cavity 1-2 phase condition by -0.14π for the other peak.

We next tune the pump powers and spatial locations for coherent interaction between all three cavities with two transparency windows, as shown in Fig. 3g. In this case, the resonant values are 1534.20 nm, 1534.35 nm and 1534.51 nm and the recorded relative delays at the EIT wavelengths are 16.48 ps and 13.29 ps. For both transparencies the detuning factor $\delta$ = 0.99. Here we calculated an extra 0.06π phase difference for the second transparency peak which we cannot remove to perfectly zero without affecting the other conditions. The estimated slow-down factors $S$ are ~350 and 115 respectively. We note that the exact spatial positions of the pumps have critical importance and allow better tuning: for instance, moving the pump that adjusts the phase, we can adjust various relative detuning of the resonant cavities without breaking the phase condition. For all of the theoretical simulations, we used $Q_v$ = 23,261 and $Q_{in}$ = 6,548 consistently as for the previous set of measurements. Longer delays can be achieved with increased number of coupled cavities and increased cavity $Q$, where $Q_{int}$ up to 100,000 or more are attainable.

In our experimental data, there is a consistent background Fabry-Perot noise due to finite reflections between the different interfaces on the chip. However, the additional EIT-like delay can be distinguished on top of the Fabry-Perot noise, and lines up well with the spectral transparency windows. The data is averaged over 64 times. For instance, when we statistically analyze the delay values and calculate the noise standard deviation $\sigma$ in our data we clearly see the difference between the Fabry-Perot noise and delay region. For example, in the two cavity interference of Fig. 3f, $\sigma$ = 5.87 ps and all the noise is between -1.99$\sigma$ and 1.76$\sigma$, whereas the maximum transparency peak shows up clearly at 2.88$\sigma$. In the three cavity interference of Fig. 3g, $\sigma$ = 5.04 ps and all the noise is between -1.48$\sigma$ and 1.92$\sigma$, whereas the maximum transparency peaks show up at 3.27$\sigma$ and 2.64$\sigma$ respectively. The transparency peak delays of 17.12 ps (Fig. 3f) and 16.48 ps (Fig. 3g) are larger (by 6.57 ps and 6.81 ps respectively) than the maximum noise, resolvable in the measurement data. The corresponding histograms for the delay values in Fig. 3f and 3g are shown in Fig. 3d and 3h respectively, where the transparency peak delay measurements are distinguished over the noise fluctuations.

In summary, we demonstrate time-domain optical delay measurements and observed slow- light in our multi-coupled photonic crystal cavities. We observe tunable delay measurements on CMOS-fabricated devices through coherent multi-cavity interactions, with delays of up to 4× the single cavity lifetime. These observations support applications towards all-optical trapping of light in a solid-state scalable implementation.

The authors thank T. Gu and J. Zheng for helpful discussions. We acknowledge funding support from 2008 NSF CAREER Award (ECCS-0747787), a 2007 DARPA Young Faculty Award (W911NF-07-1-0175), and the New York State Foundation for Science, Technology and Innovation.

**Figure 1 | Designed and fabricated L3 coupled cavity system**. **a**, Simplified model of the system which is consisting of four point-defect cavities in a 2D PC slab. **b**, Example of near-infrared top view image of 2 cavities with very close resonant frequencies. **c**, $E_y$ field intensity of coupled-cavity transparency mode between two L3 cavities. **d**, SEM of the fabricated sample with $a = 420$ nm; $r = 0.261a$; $t = 0.595a$. Each cavity is tuned ($s_1 = 0.15a$) for high intrinsic $Q$. $L_{12}=12a$ and $L_{23}=30a$. Scale bar: 5 µm. **e,** SEM of one of the cavities with higher resolution. Scale bar: 1 µm.

**Figure 2 | Temporal delay measurement setup**. The straight lines represent optical fibers and dashed lines are coaxial cables. A high speed electro-optic modulator generates a sinusoidal probe beam of 1 GHz. Two 532 nm continuous wave lasers with a 5 µm spot size at a cavity and interconnecting waveguide region are used for thermo-optic tuning. Refractive index change is $1.85 \times 10^{-4}$ / K at 300 K and we estimate 16 K temperature rise per milliwatt pump. Pump positions and powers are carefully selected to precisely control the resonant frequencies and the phase between the cavities in order to get the ideal coherent interference.

**Figure 3 | Comparison between couple mode theory calculations and the experimental results.** Measured and theoretical transmission line shapes with various detuning and the phase differences. **a.** Cavity resonances are carefully adjusted and there is no transparency window. **b** and **c.** Three resonances with spectral and phase detuning controlled for transparency peak windows. **f-g.** Measured and theoretical optical delay corresponding to 17.12 ps for **b** and 16.48 and 13.29 ps for **c**. **d** and **h.** Histograms of the delay values in **f** and **g** where the delay values at the transparency frequencies are distinctive over the noise distribution

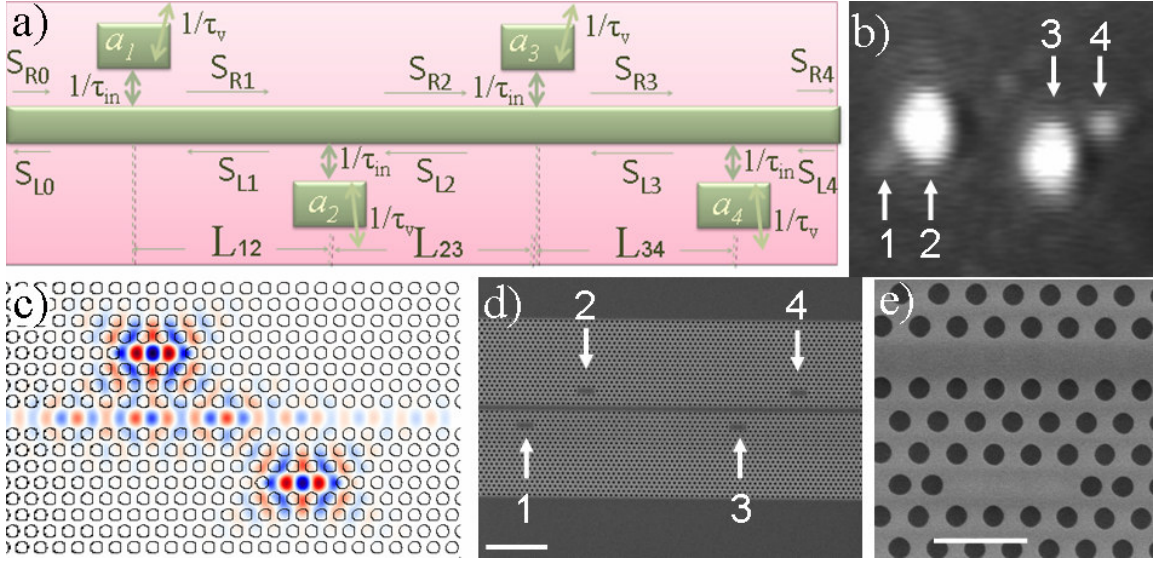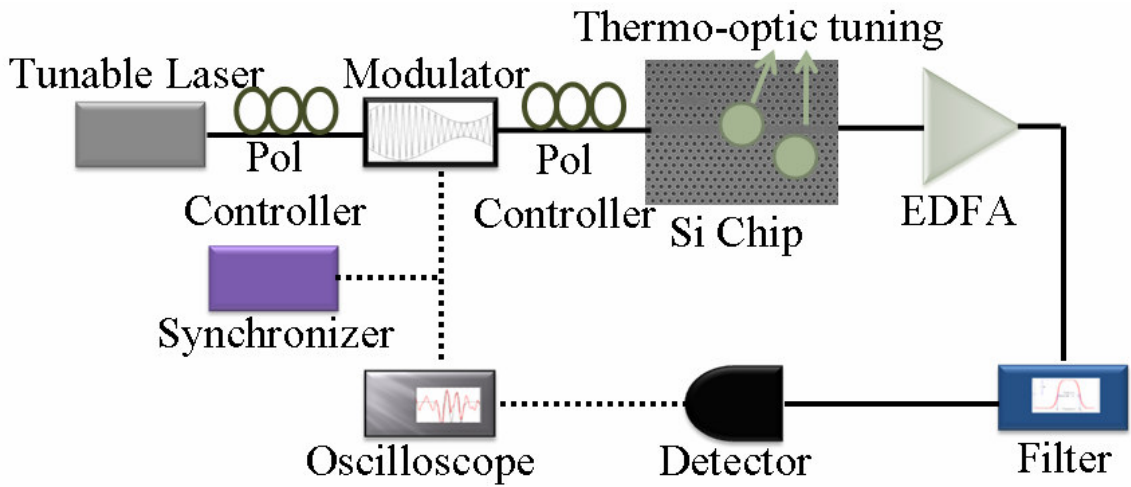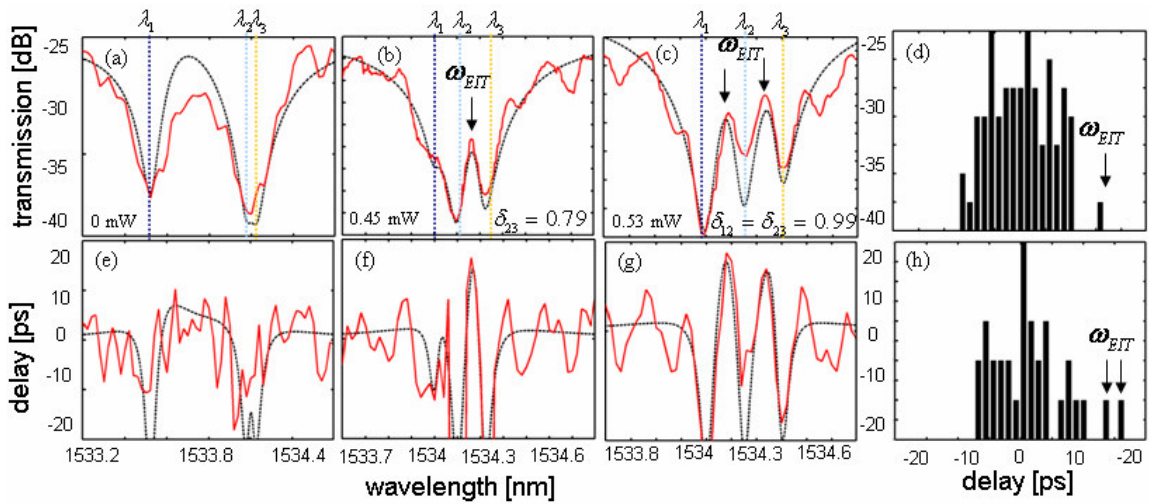